\documentclass[epj]{svjour}
\newif\ifpdf\ifx\pdfoutput\undefined\pdffalse\else\pdfoutput=1\pdftrue\fi
\newcommand{\pdfgraphics}{\ifpdf\DeclareGraphicsExtensions{.pdf,.jpg}\else\fi}
\usepackage{graphicx}
\usepackage{dcolumn}
\usepackage{bm}
\usepackage{amsmath, amsfonts, amssymb}
\usepackage{times}

\begin{document}
\pdfgraphics
\title{Charge-carrier transport properties of ultrathin Pb films}
\author{I. Vilfan\inst{1} and
 H. Pfn\"{u}r\inst{2}}
\institute{J. Stefan Institute, Jamova 39, SI-1001 Ljubljana, Slovenia, 
\email{igor.vilfan@ijs.si} \and
Institut f\"{u}r Festk\"{o}rperphysik, Universit\"{a}t Hannover,
 Appelstrasse 2, D-30167 Hannover, Germany, 
\email{pfnuer@fkp.uni-hannover.de} }
\date{Received: date / Revised version: date}
\abstract{
The charge-carrier transport properties of ultrathin metallic films
are analysed with ab-initio methods  using the density functional theory (DFT)
on free-standing single crystalline slabs in the thickness range between 1 and 
8 monolayers and compared with experiments for Pb films on Si(111).
A strong interplay between bandstructure, quantised in the direction normal to 
the ultrathin film, charge-carrier scattering mechanisms and magnetoconduction 
was found. Based on the bandstructure obtained from the DFT, we used standard 
Boltzmann transport theory in two dimensions to obtain results for the 
electronic transport properties of 2 to 8 monolayers thick Pb(111) slabs with 
and without magnetic field. 
Comparison of calculations and experiment for the thickness dependence of the
dc conductivity shows that the dominant scattering mechanism of electrons is 
diffuse elastic interface scattering for which the assumption of identical 
scattering times for all subbands and directions, used in this paper, is a good
approximation. Within this model we can explain the thickness dependences 
of the electric conductivity and of the Hall coefficient as well as the anomalous 
behaviour of the first Pb layer.
}
\titlerunning{Galvanomagnetic properties of ultrathin Pb films}
\authorrunning{Vilfan and Pfn\"{u}r}
\maketitle
               \section{Introduction}                     %
\label{intro}
Spatial confinement of the electron motion in the direction normal to the 
film causes quantisation of the perpendicular momentum and consequently a 
discrete set of electron energy subbands.
This leads to a variety of quantum-size effects 
(QSE), like oscillation of the charge-carrier density, Fermi energy, electrical 
conductivity and of the Hall coefficient as a function of film thickness, 
to mention only a few of them.
Although appreciable progress has been made in understanding the QSE,
many aspects are still unclear. 
Here we will discuss the electronic band structure and its consequences on the 
electronic charge transport with and without magnetic field in ultrathin films.

The \textit{classical} size effect on the conductivity of thin films is explained 
with the Fuchs and Sondheimer theories \cite{F38,S52,S83} which relate the 
electrical conductivity in the film and surface or interface scattering.
Their approach is based on the Boltzmann transport equation for \textit{bulk} 
electrons that are partially reflected and partially scattered on the film surfaces.
Assuming \textit{diffuse} and \textit{elastic} surface scattering they find a 
linear thickness dependence of the conductivity,
often in agreement with experiments even for film thicknesses $d$ much smaller 
than the electron elastic mean free path $\ell$, although the picture of 
3-dimensional electron motion with specular reflections at the surface or 
interface makes sense only when $d >> \ell$.
A linear $d-$dependence of the conductivity has also been obtained by Camblong 
and Levy \cite{CL99}, who used the Kubo formalism and assumed a free-electron 
model with diffuse scattering on the boundaries.

A \textit{quantum} size effect is the oscillation of the conductivity with
thickness $d$, observed in the layer-by-layer growth mode of thin films.
These oscillations have the period of the perpendicular lattice spacing and are
caused by the roughness of films with partial monolayers that are not closed.
For film thicknesses $d << \ell$ the discrete nature of the electron energy 
levels becomes essential and other quantum-size induced effects emerge on top 
of the layer-by-layer induced oscillations.
The transport of free electrons in confined geometries has been discussed by 
Sandomirskii \cite{S67} and later by Trivedi and Ashcroft \cite{TA88}.
They predicted a saw-tooth like oscillations of the in-plane conductivity with 
thickness. 
The period of the oscillation is equal to half the Fermi wavelength 
$\lambda_F$ \cite{TA88}. 
In reality, however, the film thickness increases in monolayer steps and not 
continuously as assumed in the above models.
Therefore, one  expects a range of thicknesses with large QSE amplitudes when 
$d$ is close to a multiple of $\lambda_F/2$, separated by regions with small QSE.

Apart from the theoretical attempts mentioned above, there have been several 
others which describe charge transport in thin films 
\cite{GD86,TJM86,FC89,C90,MS00,JE01,PH01}. 
Common to most of them is the assumption of free electrons. 
However, models based on free electrons break down if they are used to explain 
the galvanomagnetic phenomena (like the Hall effect), which depend on details 
of the band structure and on the electron/hole ratio, as will be shown below. 
More quantitative models are needed that take into account the detailed band 
structure of the film under size quantisation.

In this paper we address the general issue of charge-carrier scattering, 
transport and Hall effect in ultrathin metallic films.
The Hall \textit{coefficient} has the advantage that it is more sensitive to the 
QSE-mediated changes in the bandstructure and less sensitive to the 
details of the scattering mechanisms than the electrical or Hall 
\textit{conductivities}.
The model calculations will be performed on free-standing Pb(111) slabs and 
compared with experiments for  Pb on Si(111).
As substrates, semiconductors are used because they have no electron 
states at the Fermi energy to which the metal conduction electrons, responsible 
for the charge transport, could couple.
Semiconductors are also good insulators in the relevant temperature range (up to 
150 K) so that they can be considered as an insulating substrate.
Depending on substrate and annealing conditions, both crystalline and 
semi-amorphous films can be grown. 
This is important because it allows a comparison of magnetoconductive 
properties of crystalline and non-crystalline films.
For the Hall coefficient the degree of crystallinity is of minor importance. 

The bandstructure of free-standing ultrathin Pb(111) films has been calculated by 
Saalfrank \cite{S92} and more recently by Matrezanini et al.~\cite{MSL01}.
They discussed the role, QSE has on the oscillations of the Fermi energy 
$\epsilon_F$ and of the density of electron states $n_F$ at $\epsilon_F$ but 
did not consider the charge-carrier transport phenomena.
Charge transport properties at low temperatures are dominated  by scattering at 
bulk and surface imperfections.
For most ultrathin metallic films with $d < \ell$,  
scattering at the surface roughness is much stronger than scattering at 
bulk imperfections, the electrical conductivity is limited by roughness of 
the boundaries to the vacuum and to the substrate. 
The variation of the conductivity with thickness $d$ of ultrathin rough metallic 
films has been investigated by Fishman and Calecki \cite{FC89,C90}. 
In case of free electrons and uncorrelated surface height fluctuations, they 
obtain a $d^{2.1}$ thickness dependence of the ``residual'' conductivity 
$\sigma = [\rho(d) - \rho(\infty)]^{-1}$, where $\rho(d)$ is the in-plane film 
resistivity,  and a Hall coefficient $R_H$ proportional to $d$. 
None of these dependences agrees with the
experiments on ultrathin Pb films on Si(111), where an approximately linear
thickness dependence of $\sigma$ and a strongly oscillating thickness dependence 
of $R_H$ has been reported \cite{JBKL,JHB96,VHPP02}.

                     \section{The Model}                  %
We treat metallic films as free-standing, ideally ordered slabs.
The density functional theory (DFT) is applied to 
calculate the electronic band structure, density of states and Fermi lines. 
Using the Boltzmann transport equation in the relaxation time approximation
and in two dimensions we then calculate the transport properties of ultrathin 
slabs.
We will see that most of the charge-carrier transport properties as a function 
of layer thickness $d$ are already reproduced in such approach.

The basis of our charge-transport calculations are the electron band 
structures.
In a slab, the translational symmetry in the direction normal to the slab is 
broken, $k_\perp$ is no longer a good quantum number, and each band splits 
into $d$ discrete \textit{subbands}.
One can imagine these subband states as standing waves between the two 
surfaces.  
Of course, the subbands transform into a continuum of surface-projected bulk 
band states in the limit $d \to \infty$. 

The question arises, when do we have to treat the electron states as confined 
in a slab and when as bulk states.
The decisive quantity is the electron \textit{coherence length} $\ell_c$. 
If $\ell_c \ll d$, it is equivalent to the \textit{bulk mean free path} and 
is determined by bulk scattering centres.
When the coherence length exceeds the slab thickness, $\ell_c > d$, the 
electrons between the two slab boundaries behave as standing waves in the 
normal direction and as extended states in the other two directions.
Therefore, in case of ultrathin slabs, one has to distinguish between the 
electron coherence length in the direction normal to the film, $\ell_c$, 
and the electron mean free path in the plane of the film, $\ell_{mfp}$, 
which is predominantly limited by the surface roughness. 

\begin{figure}[ht]  
   \begin{center}
   \resizebox{0.5\columnwidth}{!}{%
   \includegraphics{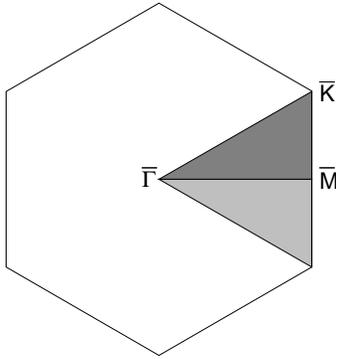}}
   \end{center}
   \caption{\label{BZ}
   Irreducible part (dark grey) of the two-dimensional hexagonal first Brillouin 
   zone of a free-standing slab.
   For the slab on a substrate the in-plane mirror symmetry is broken and the 
   irreducible Brillouin zone comprises both grey areas.
} 
\end{figure}

The DFT calculations were performed on 1 to 8 monolayers thick, 
(111)-oriented Pb slabs, separated by $\sim 10$ {\AA} of vacuum, and with 
periodic boundary conditions.
If the vacuum layer were too thin, we would observe dispersion of the electron 
bands in the $z$ direction. 
Indeed, we observed no dispersion in the occupied bands.
The electron band energies and the total energy were calculated 
\textit{ab-initio} with the full-potential linearised augmented plane-wave 
method in the local-density approximation \cite{PW92} as implemented in the 
WIEN2k code \cite{BSL01}. 
A mixed basis set of augmented plane waves plus local orbitals (APW+lo) 
\cite{S00} for low orbital momenta ($l \le 2$) and linearised augmented plane 
waves (LAPW) for all the higher orbital momenta were used.  
The spin-orbit interaction was included selfconsistently by applying the 
second-variational method and using the scalar-relativistic eigenfunctions as 
basis \cite{S94}. The Pb muffin-tin radius was set to 2.6 a.u. and a tetrahedral 
mesh of about 600 $k-$points in the irreducible part of the Brillouin zone, 
Fig.~\ref{BZ}, was used in the self-consistent electronic structure 
calculations.
The kinetic-energy  cutoff was set to $E_{\rm max}^{\rm wf} = 9.5$ Ry
and the plane-wave expansion cutoff to  $E_{\rm max}^{\rm pw} = 196$ Ry.  
Later, in the calculation of Fermi lines and conductivities, we used a mesh 
with about 4000 $k$ points in the irreducible part of the Brillouin zone.

First, the energy of bulk Pb FCC crystal was minimised to find the equilibrium 
bulk lattice constant, $a_{\rm calc}=4.88$ {\AA}, which is 1.4\% smaller than 
the experimental room-temperature value $a_{\rm exp}= 4.95$ {\AA} and 3\% 
smaller than the calculated value reported by Materzanini et al. who used the 
Cambridge Serial Total Energy Package \cite{MSL01}.  
For the slab calculations, we used the corresponding in-plane hexagonal 
lattice constant, 
$a = a_{\rm calc} / \sqrt{2} = 3.45$ {\AA}. 
Except for $d=1$ ML, no visible change in the bandstructure is observed if either 
the experimental lattice constant was used in the simulations or if the lattice 
was unrelaxed.

The transport properties are calculated using the Boltzmann transport equation 
for two-dimensional systems in the relaxation-time approximation.
The two-dimensional description is necessary when the charge-carrier mean free 
path exceeds the slab thickness. 
As we shall see later, this condition is fulfilled for all but the thinnest 
($d=1$) slabs.
For polycrystalline or amorphous metallic films the transport properties are 
\textit{isotropic} in the plane of the surface and the conductivity is 
\cite{SAT92}
\begin{eqnarray}
   \sigma_0 = \frac{ e^2}{4\pi^2\hbar d} \sum_n
    \oint {\rm d}\ell_n\; \tau_n(\epsilon_F) |\vec{v}_n(\epsilon_F)| 
    \label{sigma01}
\end{eqnarray}
where $e$ is the electron charge, $\vec{v}_n$ the group velocity of an
electron in the subband $n$ at the Fermi energy $\epsilon_F$
and $\tau_n(\epsilon_F)$ its relaxation time. 
The sum runs over all subbands crossing the Fermi energy and the integral 
is along each Fermi line. 
The spin degeneracy is included in the prefactors. 
The Hall conductivity is \cite{SAT92}
\begin{eqnarray}
   \sigma_H = \frac{ e^3}{(2\pi\hbar)^2d} \sum_n \tau_n^2(\epsilon_F)
   \oint {\rm d}\ell_n \frac{1}{|\vec{v}_n(\epsilon_F)|}\nonumber\\
   \times \left(v_x^2\frac{\partial v_y}{\partial k_y} 
    + v_y^2\frac{\partial v_x}{\partial k_x}\right).
   \label{sigmaH}
\end{eqnarray}  
 
The mechanisms influencing the relaxation time are elastic scattering on 
bulk imperfections and elastic scattering on rough surfaces, 
$1/\tau = 1/\tau_b + 1/\tau_s$. 
Inelastic scattering on phonons can be neglected if the conductivities are
investigated at low temperatures.
In ultrathin films, the main scattering mechanism is elastic scattering on 
surface roughness for which $\tau_n(\epsilon_F)$ is given by 
\begin{eqnarray}
\frac{1}{\tau_n(\epsilon_F)} = \frac{2\pi}{\hbar} \sum_{n^\prime k^\prime} 
                     |\langle n^\prime k^\prime|{\cal H}_i|n k\rangle|^2 
                     \delta(\epsilon_{n^\prime}(k^\prime) - \epsilon_F)
\end{eqnarray} 
where $ \langle n^\prime k^\prime|{\cal H}_i|n k\rangle$ is the scattering 
matrix element. 
For small, uncorrelated islands, the matrix element is independent of $n$ and $k$,
it is equal to $U^2 \rho_s /S d^2$ where $U$ is the strength and $\rho_s$ the 
surface density of the scattering centres \cite{C90}. $S$ is the area of the film.
If this situation is fulfilled, the scattering is \textit{diffuse} \cite{CL99} and 
the scattering rate simplifies to
\begin{eqnarray}
    \frac{1}{\tau} = \frac {U^2 \rho_s}{2 \pi \hbar^2 d^2}\sum_n \oint 
    \frac{{\rm d}\ell_n}{|v_n(\epsilon_F)|}.
    \label{rel_time}
\end{eqnarray} 
In this case the charge-carrier momentum is lost completely, scattering causes 
transitions  between all subbands that cross $\epsilon_F$ and also the relaxation 
time of the electrons at $\epsilon_F$ is \textit{independent} of the subband index 
$n$ and electron momentum $k$. 
In general, $v_F$ is only weakly $k-$dependent so that $\tau \propto d^2/L$ results, 
where $L$ is the length of all Fermi lines. As we shall see later, $L$ and 
therefore also $\tau$ and $\sigma_0$ are approximately linear functions of $d$.

                   \section{Results}                      %
\begin{figure}[ht] 
 \begin{center}
   \includegraphics[width=0.6\columnwidth]{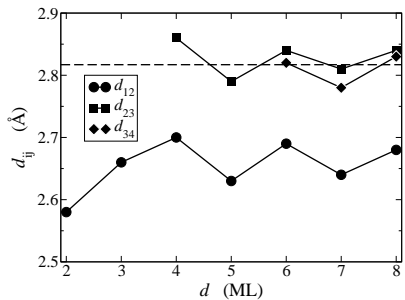}
   \end{center}
   \caption{\label{spacings}
   Interlayer spacings $d_{ij}$ (in {\AA}) of relaxed symmetric Pb(111) slabs
   as a function of slab thickness. 
   The lowest indices correspond to the surface layers.
   Dashed line indicates the relaxed bulk spacing.
} 
\end{figure}

We start our analysis by presenting the results of the lattice relaxation and 
bandstructure calculations of relaxed Pb(111) slabs. 
The thickness dependence of the interlayer spacings is shown in 
Fig.~\ref{spacings}.
The topmost layer relaxes always inwards.
In addition, we find a quantum-size induced oscillation of $d_{ij}$ as a function of 
thickness; we observe simultaneous inward relaxation of all investigated
layers for $d=2$, 5 and 7 ML thick films.
Whereas the relaxations of the topmost layers agree quite well with the 
values reported in Ref.~\cite{MSL01}, the relaxations of the inner 
layers differ. Our spacings $d_{23}$ show similar QSE oscillations as in  
Ref.~\cite{MSL01}, but they are systematically smaller 
($\sim 2$\% of $d_\textrm{bulk}$, $d_\textrm{bulk}$
is the bulk interlayer spacing in the (111) direction).
The amplitude of the QSE oscillations in $d_{34}$ is $\sim 1.5$\% of $d_\textrm{bulk}$
and is thus comparable to the amplitude of $d_{34}$ in Ref.~\cite{MSL01}, but with
opposite phases as a function of layer thickness.  

\begin{figure}[tb] 
 \begin{center}
   \includegraphics[width=0.6\columnwidth]{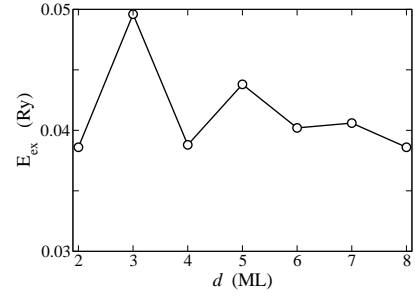}
   \end{center}
   \caption{\label{excess_en}
   Thickness dependence of the excess slab energies $E_\textrm{ex}$  
   of relaxed symmetric Pb(111) slabs. 
}
\end{figure}

Fig.~\ref{excess_en} shows the excess slab energies 
$E_{\rm ex} = E_{\rm slab} - d E_{\rm bulk}$, where $E_{\rm slab}$ is the
total energy of a slab and $E_{\rm bulk}=-41834.809$ Ry the energy/atom in 
the bulk.
Clearly seen are the QSE-induced oscillations in $E_{\rm ex}$ with minima 
for 2, 4 and presumably also 8 ML thick slabs, in partial agreement with the DFT 
calculations of Ref.~\cite{MSL01}. 

\begin{figure}[tb]   
    \begin{minipage}[t]{0.45\columnwidth}
    {\includegraphics[height=1.5\columnwidth]{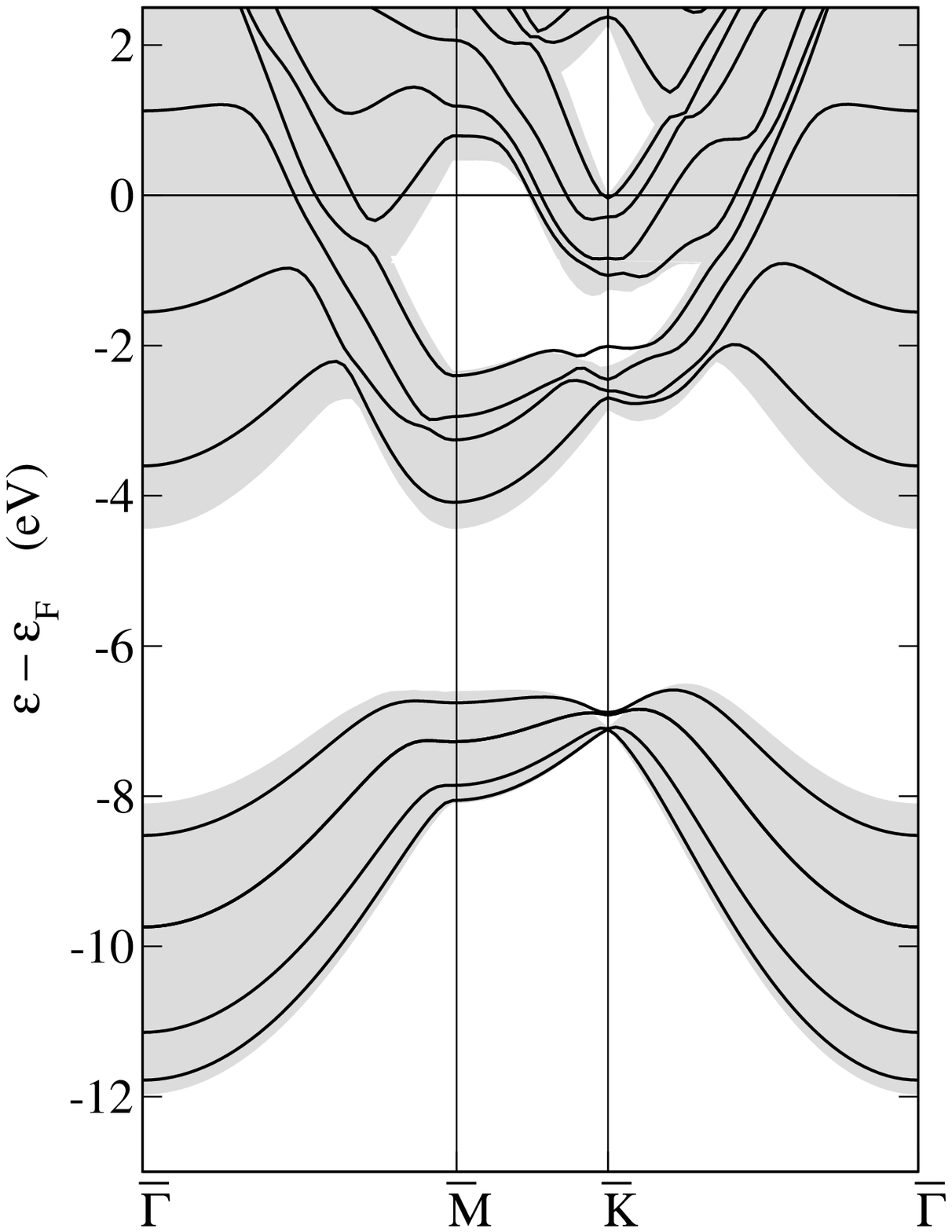}}
    \end{minipage}
    \begin{minipage}[t]{0.45\columnwidth}
    {\includegraphics[height=1.5\columnwidth]{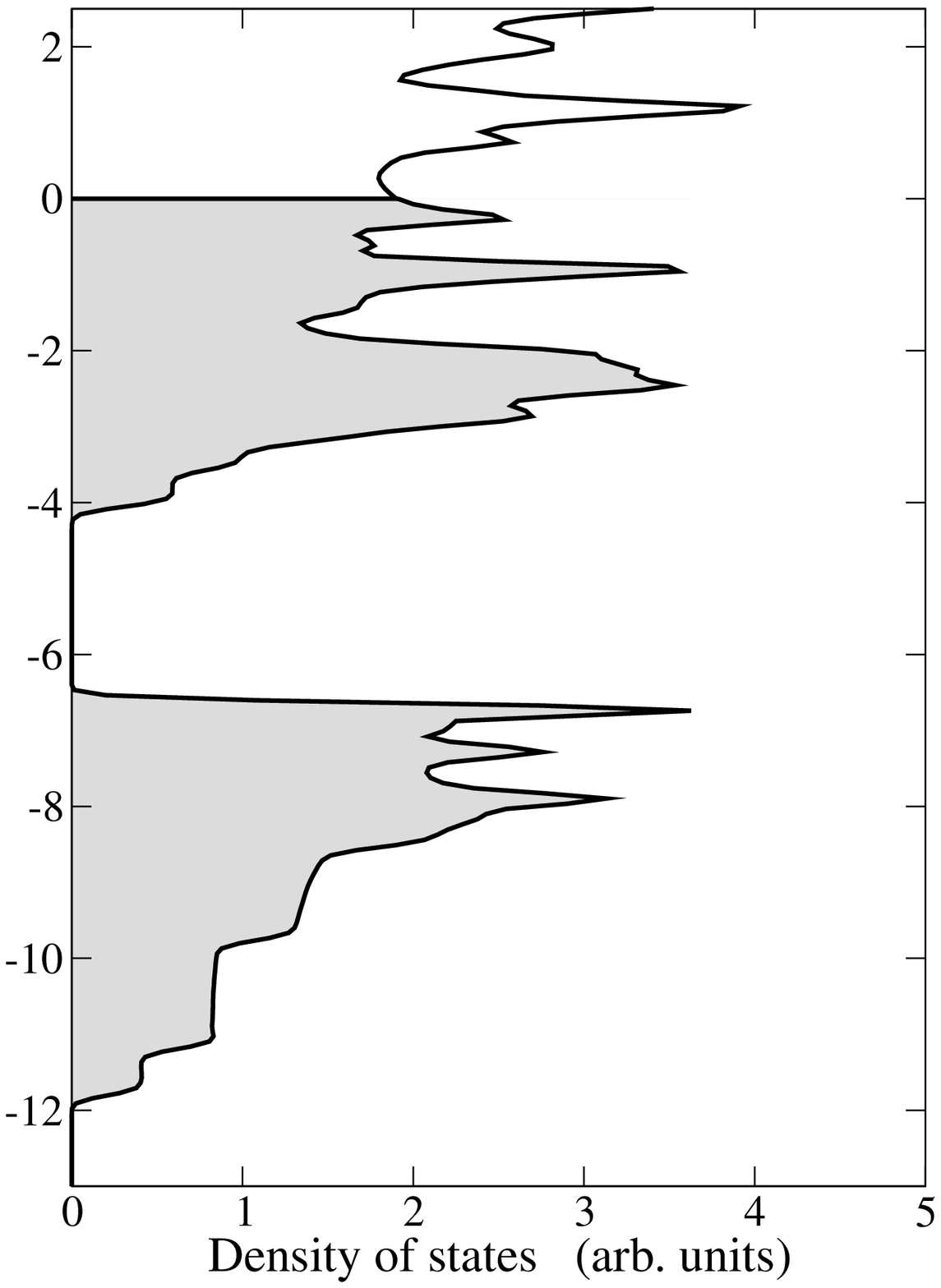}}
    \end{minipage}
    \caption{\label{4layers}
    Calculated bandstructure (left) and density of states (right) of a 
    free-standing 4 ML thick Pb(111) film.
    The shaded areas in the bandstructure represent the surface-projected bulk 
    electron energy bands, whereas the shaded areas in the density-of-states 
    are the occupied states of the 4 ML thick Pb film. 
    }
\end{figure}

\begin{figure}[tb]  
    {\includegraphics[width=0.6\columnwidth]{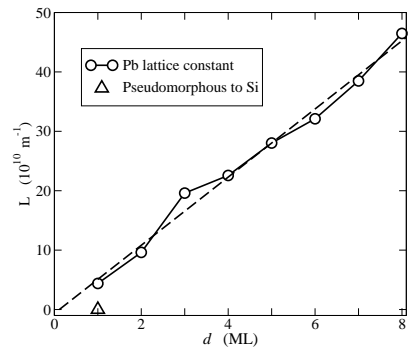}}
    \caption{\label{fermi_length}
    Thickness dependence of the total length $L$ of calculated Fermi lines
    in the two-dimensional Brillouin zone. Dashed line is a linear fit through
    the the calculated points. Also shown is the calculated Fermi-line length 
    of a $d=1$ ML thick film, pseudomorphous with the Si(111) substrate.
    } 
\end{figure}

The electron energy bands along the main symmetry directions of the 
two-dimensional hexagonal Brillouin zone for $d=4$ are shown in 
Fig.~\ref{4layers}.
Each band is split into $d$ subbands which do not cross because of spin-orbit 
interaction.
The number of subbands crossing $\epsilon_F$ is roughly proportional to $d$.
This property will have direct consequences on $\tau$ and on the conductivities.
The points of subband crossings with $\epsilon_F$ form two-dimensional closed loops,
i.e., Fermi lines.
Depending on the occupancy of the states enclosed by a Fermi line we have hole- or 
electron-like carriers. Their ratio will play a crucial role in determining
the sign of the Hall coefficient $R_H = \sigma_H/\sigma_0^2$.  
Also shown in Fig.~\ref{4layers} is the electron density of states.
Close to the bottom of the subbands the density of states has steps, 
specific for parabolic bands in two dimensions.
Proportional to $d$ is also the total length of Fermi lines, $L$, in the 
two-dimensional Brillouin zone, see Fig.~\ref{fermi_length}. This, together with
diffuse scattering, is the main reason why $\sigma_0$ is approximately proportional
to $d$. 

\begin{figure}[tb] 
    {\includegraphics[width=0.6\columnwidth]{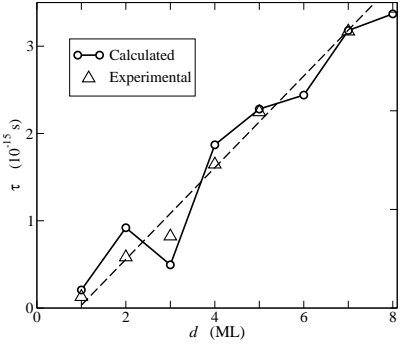}}
    \caption{\label{tau}
    Thickness dependence of the charge-carrier scattering relaxation time 
    $\tau$ assuming equal roughness (equal $U^2\rho_s$) of all slab 
    thicknesses $d$.
    In case of diffuse elastic scattering, $\tau$ is independent of the 
    subband index $n$ and electron momentum $\vec{k}$.
    Experimental points are given by Eq.~(\ref{tau_exp}) where the experimental
    conductivities $\sigma_0^\textrm{exp}$ are from Ref.~\cite{VHPP02}.  
    Dashed line is a linear fit to the experimental points.}
\end{figure}

\begin{figure}[tb]   
     \includegraphics[width=0.6\columnwidth]{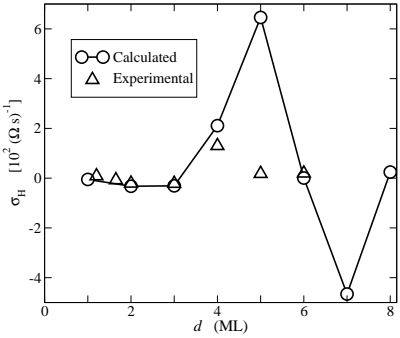}
     \caption{\label{sigH}
     The calculated Hall conductivity $\sigma_H$,
     Eq.~(\ref{sigmaH}), is compared to the experimental Hall
     conductivity, $\sigma_H^\textrm{exp} = R_H (\sigma_0^\textrm{exp})^2$.
     Huge oscillations are the effect of cancellations between electron and 
     hole contributions.
     }
\end{figure}

We next calculated the relaxation time, Eq.~\ref{rel_time}, assuming 
a potential depth 3 eV, a terrace height one ML ($\sim 3$ \AA),
a terrace area 100 \AA$^2$  ($U=900$ eV\AA$^3$) and a terrace
density $\rho_s = 10^{-4}$ \AA$^{-2}$,
see Fig.~\ref{tau}. These numbers were chosen in order to simulate closely
the experimental situation described in Ref.~\cite{Pfen02} and to obtain a close
fit to the experimental values for $\sigma_0$ and $R_H$ \cite{Pfen02}.
We note that elastic scattering at the surfaces causes transitions between 
\textit{all} subbands that cross the Fermi energy.
The summation over all final states in the scattering rate (\ref{rel_time})
causes $\tau \propto d$. As we see from Fig.~\ref{tau}, $\tau \propto d-1$
describes the thickness dependence better. The reason is found in the details of
the bandstructure (see below).  
The deviations from linearity at $d=3$, 6 and 8 ML appear when the Fermi energy 
coincides with a local peak in the electron density of states, thus opening more 
channels for elastic scattering.
Also shown in Fig.~\ref{tau} is the ``experimental'' value $\tau^\textrm{exp}$,
\begin{eqnarray}
   \tau^\textrm{exp} = \frac{4 \pi^2 \hbar d\; \sigma_0^\textrm{exp}}
    {e^2 \oint dl |v_n(\epsilon_F)|},
    \label{tau_exp}
\end{eqnarray} 
obtained from Eq.~(\ref{sigma01}) by combining $\sigma_0^\textrm{exp}$ from 
Ref.~\cite{VHPP02} with the calculated value for $\oint dl |v_n(\epsilon_F)|$.
We also calculated $\sigma_0$, Fig.~\ref{sig0}, and 
$\sigma_H$, Fig.~\ref{sigH}.
$\sigma_0$ has a very similar thickness dependence as $\tau$, indicating,
that  $\tau$ is the main source of the thickness dependence of $\sigma_0$.

The Hall conductivity, $\sigma_H$, on the other hand, is a complicated
function of $d$, mainly because of compensating effect of electron and
hole subbands contributions. 

\begin{figure}[tb]  
    \includegraphics[width=0.8\columnwidth]{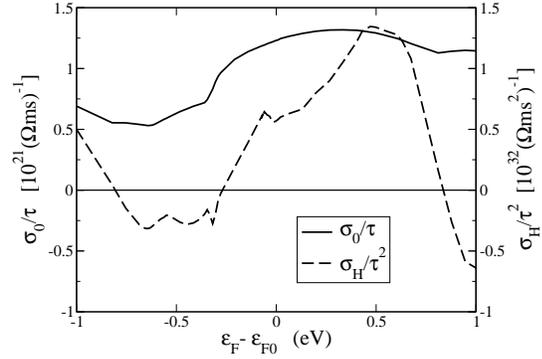}
    \caption{\label{sigma0H}
    Energy dependence of the calculated electrical conductivity 
    $\sigma_0/\tau$ 
    and Hall conductivity $\sigma_H/\tau^2$ for a 4 ML thick slab.
     }
\end{figure}

There have been several attempts to relate the Hall coefficient to the electron
density of states, in particular for disordered metals \cite{HMDO00}. 
Therefore we decided to test the above relations. 
In Fig. \ref{sigma0H} we show the dependence of calculated $\sigma_0/\tau$ and 
$\sigma_H/\tau^2$ on the position of the Fermi level for a $d=4$ ML slab. 
We artificially shift the Fermi energy and calculate the corresponding 
$\sigma_0/\tau(\epsilon_F)$ and $\sigma_H/\tau^2(\epsilon_F)$. 
As is seen from this figure,   
$\sigma_0/\tau$ is a relatively smooth function of $\epsilon_F$ whereas
$\sigma_H/\tau^2$ oscillates strongly with $\epsilon_F$
and changes sign:
it is negative at the bottom of each subband, when the charge carriers are 
electron-like, and positive at the top of the subbands when the carriers are 
hole-like. 
In between, they are complicated functions of the Fermi energy and reflect 
different singularities in the bandstructure.
As a consequence, the dependence of $R_H$ on $\epsilon_F$ is given 
predominantly by $\sigma_H/\tau^2$.
Comparison  of $\sigma_H/\tau^2(\epsilon_F)$, Fig.~\ref{sigma0H}, and of the 
density of states, Fig.~\ref{4layers}(b), 
shows that for $d=4$ ML, there is no evident relation between $R_H$ and the 
density of states.
Such a relation would be valid only if we had a single three-dimensional band, 
i.e. when the charge carriers at the bottom (top) of the band are electrons 
with $R_H <0$ (holes with $R_H > 0$) and at the same time the density of 
states is a rising (falling) function of the energy. 
In case of several inter-penetrating bands, as is the case of thin films, such a 
simple picture breaks down \cite{BG74,M85,NMMP87}, as is also obvious from our 
results. We also find no obvious relation between $\sigma_0$ and the density of 
states.

     \section{Comparison with the experiments}            %
\begin{figure}[tb]  
     \includegraphics[width=0.6\columnwidth]{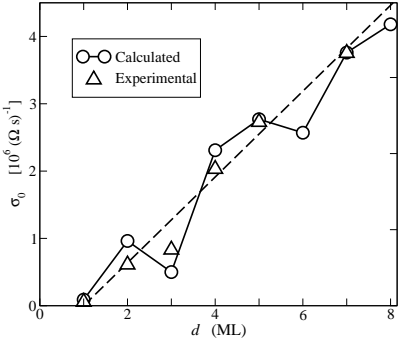}
     \caption{\label{sig0}
     Thickness dependence of the calculated [Eq.~(\ref{sigma01})] and 
     low-temperature experimental \cite{VHPP02}  conductivity $\sigma_0$.
     Notice the remarkable similarity between $\sigma_0$ and $\tau$ which
     is the consequence of linear thickness dependence of $L$, Fig.~\ref{tau}.
     Dashed line is a linear fit through the experimental points.
     (In Fig.~2 of Ref.~\cite{VHPP02} $\sigma_0$ must be multiplied
     by a factor 100.)
     }
\end{figure}

In Figs.~\ref{tau} and \ref{sig0} we compare calculated $\tau$ and $\sigma_0$
with the experiments.
It has been observed experimentally that the transport properties of Pb films 
are almost independent of the substrate reconstruction \cite{VHPP02}. 
Therefore, metals on Si can be considered to a reasonable first approximation as 
free-standing metal films. This is not true for the first Pb layer, 
which will be discussed separately below. 
 
Clearly seen in theory and experiments is an approximately linear 
thickness dependence of $\sigma_0$ which corroborates the picture
that the scattering is dominated by surface roughness and that 
the roughness does not change with film thickness.

We identify several conductivity regions.
Below the percolation threshold at $d \sim 0.8$ ML the charge carriers are 
localised, the transport is strongly temperature activated and $\sigma_0$ 
is so small that it is of no relevance for our considerations.

Between the percolation threshold and 1 ML the Pb films are pseudomorphous, 
i.e. the lattice constant is $\sim 10\%$ larger than the corresponding bulk 
Pb value.
We performed also DFT and conductivity calculations for a $d=1$ ML thick Pb
slab with the lateral lattice constant of a Si(111) substrate and we find that
the electrical properties of a 1 ML film are extremely sensitive to
the lateral lattice constant $a$. 
A 10\% increase in the lattice constant changes the metal to an indirect bandgap
semiconductor. 
The experimental conductivity of a 1 ML film is very small and has an approximately
linear temperature dependence \cite{VHPP02}, supporting the theoretical conjecture 
that the 1 ML Pb(111) films are semiconducting. 

Between $\sim 1$~ML and $\sim 3.5$~ML, the annealed films are amorphous  and 
$\sigma_0^{\rm exp}$ exhibits a small, almost linear increase with 
temperature up to $\sim 100$ K, indicating a small activated contribution. 
Films with $d \gtrsim 4$~ML  are crystalline and $\sigma_0^{\rm exp}$ 
decreases with increasing temperature (phonons).
We could not observe any influence of disorder on the conductivity of 
ultrathin films. The experimentally observed dip in
$\sigma_0$ at 3 ML (see Fig.~\ref{sig0}) is fully reproduced in our calculation 
with ideally ordered slabs. 
It is related to the electronic bandstructure, but has nothing 
to do with disorder in the film. Therefore, it is a pure quantum-size effect.

The thickness dependence of $\sigma_0$ is, like $\tau$, $\propto d-1$. 
It is governed by scattering on the surface or interface roughness --  
bulk disorder would give a thickness-independent $\sigma_0$.
The main source of the oscillations of $\sigma_0$
with $d$ is the quantisation of the Pb band structure in the direction normal 
to the film together with the strong spin-orbit interaction. 

Our results for $\sigma_0$ are consistent with the picture that the 
electrons are subject to diffuse elastic scattering at the surfaces of the 
films \cite{F38,S52,CL99}, in agreement with the original Fuchs-Sondheimer 
model, but they are at variance with the model of scattering at surfaces with 
uncorrelated roughness, where $\sigma\propto d^2$ was predicted \cite{C90}.

The calculated Hall conductivity is compared to $\sigma_H^\textrm{exp}$
in Fig.~\ref{sigH}.
The agreement is good except for $d=5$ ML where the discrepancy most probably
lies in the lifetime broadening of the electron states \cite{VHPP02} which is
neglected in the present analysis.

                   \section{Discussion}                   %
     
Our calculation used for the interpretation of the experimental data  is based 
on a DFT band structure calculation combined with the Boltzmann transport 
equation. 
The only adjustable parameter, used in the calculation, is the scattering
strength $U^2 \rho_s$.
The physical scenario is described by electronic states in ultrathin films 
that are localised in the direction normal to the surface and are scattered
by surface roughness. 
The electron states form discrete subbands since the electron coherence length,
limited  by scattering in the film bulk, exceeds the film thickness.

The results of the calculation are consistent with the picture that the 
electrons are subject to diffuse elastic scattering on film surfaces
\cite{F38,S52,CL99} for which $\sigma_0$ and $\tau$ 
are proportional to the number of conducting metallic layers, in our case 
$(d-1)$, in agreement with the original Fuchs-Sondheimer model and with the 
experiments \cite{VHPP02}.
The (lateral) mean free path of electrons in Pb ultrathin films on Si(111) exceeds 
the Pb bulk lattice constant when $d > 1$ ML. 
According to the Ioffe-Regel criterion \cite{IR60}, $l \sim 1/k_F \sim a_b$
($a_b$ is the bulk lattice constant), the electrons in  $d>1$ ML thick films are 
not localised, and the use of the Boltzmann transport equation is justified.

The essential difference between the Fuchs-Sondheimer and the present model is
that in the former the electrons are described by three-dimensional Bloch states 
whereas in our approach the electron states are described by two-dimensional 
momenta $k_\|$ and discrete subbands. The former approach is adequate for thick 
slabs, whereas our approach is valid for ultrathin slabs when the electron 
coherence length exceeds the slab thickness $d$.

In experiment, no free-standing films can be investigated, i.e. an insulating 
substrate is always required that tries to impose its own periodicity on the 
adsorbed film, depending on the interface energies, and on lattice mismatch. 
For a perfect film, an atomically sharp interface and a well defined 
periodicity at the interface are required (e.g., in form of a superstructure). 
In the Pb/Si(111) case, where the lattice mismatch is approximately 10\%, no 
lattice match, apart from the first monolayer, can be expected. 
This is the reason for the formation of the amorphous films up to 4 monolayers,
whose many degrees of freedom are obviously needed to allow for the growth of 
a Pb film with its own lattice constant for thicker layers. 
Nevertheless, it seems that the short range order in these amorphous
films is still sufficiently good to allow a comparison of conductivity and 
Hall conductivity calculated for perfect crystalline order with experimental 
data obtained in amorphous films. The calculations with perfectly ordered 
films even reproduce details of the conductivity as a function of layer thickness.

As seen from our analysis, the diffuse interface scattering mechanism 
still limits the conductivity, even for the crystalline films at larger $d$. 
For crystalline layers the interface to the vacuum is atomically smooth for 
complete layers, but can be roughened by adsorption of incomplete layers, as 
seen by a reduction in conductivity \cite{Pfen02}. 
These roughness-induced changes of conductivity, however, are small. 
Therefore, the Pb-vacuum interface does not seem to provide the main contribution 
to diffuse scattering for the crystalline Pb films (the terraces are too large
and their density too small).  
An electronically still very rough interface, on the other hand, may exist 
between the crystalline and the amorphous layers. 
Because of the lack of order in the amorphous state, this interface scatters at 
random in all directions, i.e. it still acts as diffusely scattering interface. 
Additional contributions to electron scattering from the Si-Pb interface are 
conceivable. 

The single pseudomorphous monolayer of Pb on Si(111) has 
a narrow indirect band gap and is thus semiconducting, according to the DFT 
calculations, in agreement with the
experiments, where the conductivity has an approximately linear temperature
dependence \cite{VHPP02}. 
This is not the case for free-standing Pb monolayers, which correspond to Pb
deposited on semiconducting 1 ML Ag-covered
Si(111)$(\sqrt{3}\times \sqrt{3})$ \cite{JHB96}, where Pb still has a Fermi 
line and is thus, strictly speaking, a metal, but with a very small conductivity.

\section{Conclusions} 

Our  analysis of the charge-carrier scattering and transport in ultrathin 
metallic Pb films has shown that they are very sensitive to the reduced 
dimensionality of ultrathin layers resulting in classical and quantum size 
effects. 
The relaxation time $\tau$ and the conductivity $\sigma_0$ are limited
by the scattering on rough surfaces and are approximately linear functions
of $d$ (classical size effect) with quantum-size induced deviations from 
linearity at $d=3$, 6 and 8 ML.
The proportionality constant in $\sigma_0 \propto (d-1)$  depends on roughness 
of the interface to the substrate.
The deviations for $d=3,6$ and 8 ML appear when the Fermi energy coincides with 
a local maximum in the electron density of states and causes a drop in $\tau$.
These deviations are the analogue of the condition  $id \approx j \lambda_F/2$
(where $i$ and $j$ are integers and $\lambda_F$ the Fermi wave length) 
in the free electron model \cite{S67,TA88}.
$R_H$ is very sensitive to details of the bandstructure and to a large extent 
insensitive to disorder in the film and to surface roughness. 

As demonstrated in this paper, the charge-carrier transport is very suitable for
investigating the QSE, since it is to a large extent insensitive to disorder in
the film and is limited by the interface roughness.

The quantum-size-induced oscillations of $\sigma_H$ and $R_H$ 
are strong and originate from partially compensating contributions to the Hall 
conductivity that emerge from different subbands crossing the Fermi energy.
The advantage of $R_H$ over $\sigma_H$ is that $R_H$ is almost insensitive to the 
charge-carrier lifetime and consequently to ordering in the film. 

Whereas one ML of free-standing Pb slab is metallic, 
a single monolayer of Pb, pseudomorphous with Si(111), is semiconducting.
In both cases, however, the conductivity and the mean free path are small. 
The band gap in the pseudomorphous case is caused by spin-orbit splitting.

The method applied here can be used for other ultrathin metallic films,
provided the electron coherence length (or mean free path) is larger than the 
film thickness.
We have shown that the quantum-size effects depend on the details of the
electronic bandstructure and cannot be treated quantitatively with the free
electron models, in particular in case of heavier metals when the spin-orbit
interaction is strong. This applies also to $\sigma_H$ which is very 
sensitive to details of the bandstructure.

\begin{acknowledgement}
This work was supported by grants from the Deutsche Forschungsgemeinschaft, 
from the Deutsche Akademische Austauschdienst and from the Slovenian MSZS.
\end{acknowledgement}



\end{document}